# Channel resolution enhancement through scalability of nano/micro-scale thickness and width of SU-8 polymer based optical channels using UV lithography

Iraj S Amiri[1,2], Volker J. Sorger[5], M Ariannejad[3], M Ghasemi[3], and P Yupapin[1,4*]

[1]Department for Management of Science and Technology Development, Ton Duc Thang University, Ho Chi Minh City, Vietnam.
[2]Faculty of Applied Sciences, Ton Duc Thang University, Ho Chi Minh City, Vietnam.
Email: irajsadeghamiri@tdt.edu.vn
[3]Photonics Research Centre, University of Malaya, 50603 Kuala Lumpur, Malaysia;
[4]Faculty of Electrical & Electronics Engineering, Ton Duc Thang University, District 7, Ho Chi Minh City, Vietnam;
[5]Department of Electrical and Computer Engineering, The George Washington University, Washington, D.C. 20052, USA

**Abstract:** This paper reports on an approach for fabrication of micro-channels with nanometer thickness achieved by optimization of UV lithography processes. Rectangular micro-channels with a staple edge are fabricated over the surface of a silicon wafer substrate in which a sub-micron layer of diluted SU-8 thin film has been coated. The optimization of the process parameters including the duration of a two-step pre- and post-baking process, UV exposure dosage, and finally chemical developing time with constant agitation produces micro-channels with high contrast edges and thickness below 100 nm is achieved. The dimensions achieved using this approach has potential applications in sub-micron optical waveguides and nanoelectromechanical (NEMS) devices.

## 1. Introduction

Soft lithography has been an interest of late especially as a different approach to rapid prototyping of both micro-scale and nano-scale structures [1]. This approach provides structure formation on planar, curved, flexible and soft substrates especially when the inexpensive method is required [2]. Normally, an alternative to photolithography is the e-beam lithography (EBL) which can produce pattern structures with feature sizes larger than 1 µm [3] down to a lateral dimension in the region of 20–30 nm [4, 5]. This is much smaller than those achievable by photolithography [6]. Currently, a number of computer-aided design (CAD) software programs are available to provide the design of patterns [7]. In the case of photolithography, UV exposure allows transfer of masks pattern onto a photoresist layer. If a thick photoresist layer (larger than 1 µm) is used, the aspect ratio must be high with good sidewall quality and also good dimensional control in order to fulfill the requirements for applications in MEMS and MOEMs [8, 9]. Ultra-thick photoresist layers are also used for fabricating electrostatic sensors, actuators and microfluidic channels [10]. Specific protocols have to be closely followed such as the optimization of spin coatings, UV exposure and so on as to achieve the best results [6]. The other important steps will be a post-baking process of the samples, in this case, the SU-8 polymer, which has been clearly explained in ref [7]. The next step will be the development of this pattern, which is achieved by immersion in the SU-8 developer with the aid of magnetic agitation to remove the unexposed SU-8 region. SU-8 has a low optical absorption in the near-UV range (320-420nm), thus allowing layer patterning with a thickness of up to a few hundred micrometers while still giving a very high aspect ratios [11]. Chang and Kim obtained 25 µm tall structures with 1µm feature size. Ling et al. obtained 360µm tall structures with 14µm feature sizes [12, 13]. A number of solvents, including SU-8 developer, ethyl lactate, and diacetone alcohol work well with SU-8 [14]. Under partial exposure, when the reflection between SU-8 and the substrate surface is significant, the thickness of SU-8 structures after development will be affected [15].

In the case of an ideal contact exposure, the mask and the photoresist surface should be as close as possible with a minimum air gap in-between. In most cases, the air gap is about 10–100 µm for thick photoresist [16]. Another concern will be the thickness of SU-8, as the thickness increases, the non-uniformity of the photoresist can become a serious issue [17], resulting the error in the surface flatness causing air gaps between the mask and the resist surface. Due to this, there will be the occurrence of Fresnel diffraction, and formation of micro patterns on the sample, thereby reducing the sidewall quality of the patterned microstructures [18]. This effect can be severe as the thickness of the photoresist layer and mask photoresist gap increases. An alternative to this will be to explore the technique of soft lithography, which offers the opportunity of a broader range of materials, as well as experimental simplicity and flexibility in forming certain types of test patterns [19].

## 2. Micro-channels Fabrication

In this section, we study the effects owning to the variation of fabrication parameters such as different SU-8 solid content, spin coating recipes, pre- and post-soft baking time, and UV exposure dosage on the dimensions of the produced micro-channels. It must be noted that, for obtaining the maximum reliability of the results, the substrates used in this work were cleaned using standard cleaning procedures. However, the major focus of this work is the thickness scalability of the micro-channels (with micron-scale width). Other parameters such as the uniformity, the quality and the loss of the channel are not considered for the current debate. Table 2 summarizes the optimization process carried out in this work. The negative tone SU-8-2002 solution, which is the improved formulation of SU-8, is diluted with cyclopentanone, a common solvent for SU-8-2000 series [20]. Based on the information provided by the data sheet of the SU-8-2000 series, the film thickness of 0.5 to > 200 microns can be obtained with a single coating process.

In spite of this range of the thickness, further reduction of the thickness to less than 0.5 microns is possible by dilution the SU-8-2000 solution with a different volume of the solvent. Hence, in our experiment, the viscosity of the SU-8-2000 polymer reduces when it is diluted with a different volume ratio of cyclopentanone, viz. 20%, 25%, 40% and 50%. The results in this section will show that the dilution is not the only factor influencing the thickness of the micro-channels, and the other factors such as the defined recipe for spin coater, pre- and post-baking time, exposure time and development time effectively alter the thickness of the micro-channels. In the following, we present the fabrication process and parameters of the SU-8 based micro-channels. Silicon substrates with a dimension of 0.5×0.5 cm$^2$ were used in this work. The standard cleaning process which is used to clean the silicon substrate consists of three steps: 1. washing the silicon substrate with acetone 2, then washing with the iso-propyl alcohol (IPA) and 3. drying the silicon substrate with nitrogen gas for around 30 seconds. The complete dehydration of the sample can be achieved by placing it on the hot plate for around 4 to 5 minutes. In the next step, we dispensed 0.5 ml of the SU-8 over the surface of the silicon substrate before spin coating commence.

Different spin-coating recipes (i.e. each recipe includes a set of time-based programs to allow the spin coater reach to the maximum speed smoothly) have been used to achieve different film thickness with sufficiently good thickness uniformity. Recipe 1 and 2 shown in Figure 1 are the illustrative examples of the ramp-up and final spin cycles corresponding to the test number 3 and 4 shown in Table 2. They reach the maximum spin speed of 5000 and 4000 RPM, respectively, within the span of several tenths of seconds. The physical observation of the coated samples according to Figure 3 and Figure 4 shows the uniformity of the photoresist layer over the silicon, viz. surface profiler has been used to evaluate the uniformity of the SU-8 samples before conducting the optical lithography.

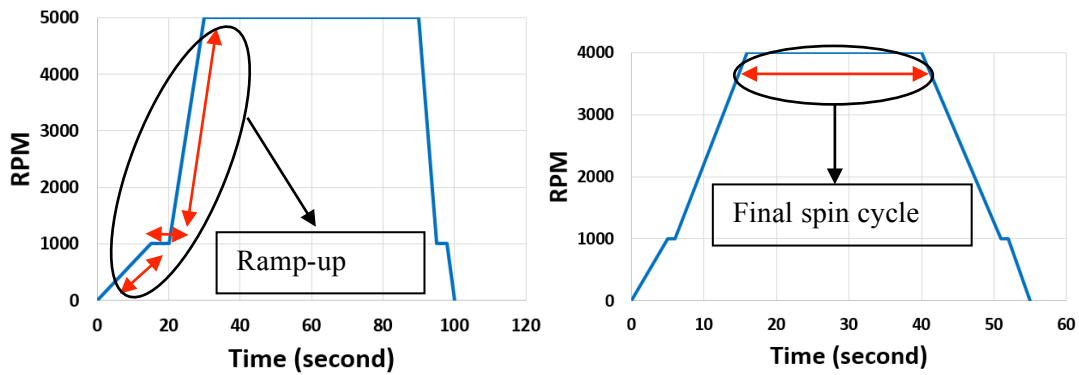

**Fig. 1:** Illustrative recipes of spin coater with different spread cycles and the final spin cycle of 5000 (a) and 4000 (b).

After the photoresist has been coated over the substrate surface, the samples undergo pre- and post-soft baking with a time span of 1 or 2 minutes (as shown in Table 2). Sample #1-6 and #13-17 Table 2 were subjected to a one-step thermal treatment of 65 °C (pre-soft baking), while samples #7-12 in Table 2, were subjected to an additional thermal treatment at a higher temperature of 95 °C (post-soft baking). Lower initial bake temperatures allow the solvent to evaporate out of the film at a more controlled rate, which results in better coating fidelity, reduced edge bead, and better resist-to-substrate adhesion. It must be noted that after each pre- and post-soft baking, the temperature of the sample need to be decreased gradually to room temperature. This will minimize the possibility of crack formation as a result of temperature-induced stress. After the soft baking process, the samples were loaded on the stage of the UV mask aligner. Lateral alignment of the sample with photomask was performed before decreasing the gap and vacuuming the air between the sample and photomask. Better contact of the photomask with the sample will enhance the quality of the printed patterns after UV exposure.

Therefore, for the sake of the accuracy of the achieved results, the quality of the contact between the sample and photomask has been checked by the aid of microscope before each time UV exposure. The SU-8 samples were then exposed to UV irradiation (365 nm) for a controlled duration. Since there is a direct relation between the photolithography parameters such as exposure dosage, wavelength, exposure time and chemical development time, for obtaining micro-channels with the desired micrometer width and nanometer thickness, these parameters were experimentally tested viz. as described in Table 2. In the next step, the exposed samples were baked again. This step which is called post exposure bake (or post baking after exposure as shown in Table 2) is the essential part of the experiment, since the thermoplastic property of the photoresist SU-8 allows to selectively formation of cross-linking between monomers of the SU-8, and as a result, polymerization of the SU-8. Crosslinking phenomenon takes place when different parts of monomer chains become covalently bound. The SU-8 molecules create the cross-links between its monomers, and consequently, this restricts the movement of the molecules of the SU-8. Increasing the number of crosslinks makes the polymer more rigid but brittle. Also, the created polymer chains cannot slip pass each other easily and thus the elasticity of such a rigid structure become highly dependent upon the number of crosslinks in the polymers.

The amount of the post-exposure time fundamentally alters the number of crosslinks and accordingly the thickness of the SU-8. More crosslinks produce thicker SU-8 patterns. SU-8, which is known as a readily cross-linked photoresist with one-time short post-baking (i.e. ~ 1 minute), produces a highly stressed film with a thinner thickness of the pattern. Thus, for solving this problem and minimizing the stress, wafer bowing and resist cracking, two step post exposure baking (at 65 °C and at 95 °C, as recommended by the manufacturer) process were performed as shown in Table 2. It must be noted that slow cooling of the sample to temperature room is performed again after each post exposure baking. The solvent used in this experiment is Ethyl Lactate. The strength of the agitation of the SU-8 sample during chemical development as an effective parameter on thickness control of the channels is also considered in this study. During the chemical development process, the SU-8 samples must be treated with the shortest immersion duration at a controlled agitation, thus megasonic device alongside a timer has been used in each chemical development process in Table 2. By defining the ratio of UV dosage to developer influence (i.e. development time and frequency of the vibration) as $R/R_0$, the cross-sectional profiles as a result of different dissolution rate after immersion of the sample in the chemical developer can take the patterns like the ones shown in Table 1. This study aimed to fabricate the planar micro-channels with a staple edge, therefore, $R/R_0$ has been selected as the control parameter.

The SU-8 samples after chemical development process have been rinsed completely with IPA, and dried with a gentle stream of nitrogen gas. The thickness, width and the quality of the channels on the processed samples after complete UV photolithography are examined using a surface profiler (DEKTAK D120) and microscope (ZEISS xxx). The surface quantification of analyzed samples is summarized in Table 2 and Figures 2-12, respectively.

**Table 1.** The relationship between the produced pattern profile of the channel, UV dosage, and chemical development parameters.

| Profile | UV light Dosage | Developer Influence | $R/R_0$ | Application |
|---|---|---|---|---|
| 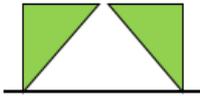 | High | Low | >10 | Lift-off, ion implant |
| 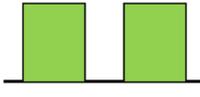 | Moderate | Moderate | 5-10 | Dry Etch, Lift-off, Wet Etch |
| 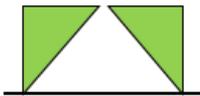 | Low | Low | <5 | Wet Etch |

**Table 2.** Fabrication parameters used for characterization of SU-8 based optical micro-channel through UV lithography

| Test number (#) | Photoresist dilution (Su8-2002) | Volume on substrate (μ lit) | Spin cycle Speed (RPM) | Pre soft bake time (min) 65°c | Pre soft bake time (min) 95°c | Exposure time (second) | Exposure dosage (mw/cm²) | Post soft bake(min) 95°c | Post soft bake(min) 65°c | Develop (second) | Megasonic frequency (MHz) | Thickness of Channel 5 (nm) | Thickness of Channel 4 (nm) | Thickness of Channel 3 (nm) | Thickness of Channel 2 (nm) | Thickness of Channel 1 (nm) | Width of Channel 5 (nm) | Width of Channel 4 (nm) | Width of Channel 3 (nm) | Width of Channel 2 (nm) | Width of Channel 1 (nm) |
|---|---|---|---|---|---|---|---|---|---|---|---|---|---|---|---|---|---|---|---|---|---|
| 1 | 20% | 500 | 3000 | 1 | 1 | 12 | 25.5 | 1 | 1 | 14 | 2 | 395 | 398 | 390 | 385 | 353 | 4870 | 3770 | 2930 | 1670 | 880 |
| 2 | 20% | 500 | 3500 | 1 | 1 | 12 | 25.5 | 1 | 1 | 14 | 2 | 177 | 164 | 155 | 150 | 141 | 4860 | 3810 | 2920 | 1740 | 910 |
| 3 | 20% | 500 | 4000 | 1 | 1 | 12 | 25.5 | 1 | 1 | 14 | 2 | 168 | 158 | 160 | 155 | 150 | 4930 | 3780 | 2690 | 1700 | 760 |
| 4 | 20% | 500 | 5000 | 1 | 1 | 12 | 25.5 | 1 | 1 | 14 | 2 | 62 | 68 | 66 | 68 | 64 | 3900 | 3800 | 2920 | 1910 | 990 |
| 5 | 25% | 500 | 4000 | 1 | 1 | 12 | 25.5 | 1 | 1 | 14 | 2 | 318 | 290 | 283 | 267 | 215 | 4954 | 3940 | 3800 | 3000 | 500 |
| 6 | 25% | 500 | 5000 | 1 | 1 | 12 | 25.5 | 1 | 1 | 14 | 2 | 372 | 365 | 355 | 338 | 311 | 4630 | 3847 | 2780 | 1777 | 840 |
| 7 | 25% | 500 | 3000 | 2 | * | 20 | 22 | 3 | * | 17 | * | 2730 | – | – | – | – | – | – | – | – | – |
| 8 | 25% | 500 | 5000 | 1 | * | 20 | 22 | 2 | * | 13 | * | 3200 | – | – | – | – | – | – | – | – | – |
| 9 | 25% | 500 | 5000 | 1 | * | 20 | 22 | 2 | * | 28 | * | 3400 | – | – | – | – | – | – | – | – | – |
| 10 | 25% | 500 | 5000 | 1 | * | 10 | 22 | 2 | * | 10 | * | 1000 | – | – | – | – | – | – | – | – | – |
| 11 | 25% | 500 | 5000 | 1 | * | 7.5 | 22 | 3 | * | 18 | * | 1600 | – | – | – | – | – | – | – | – | – |
| 12 | 25% | 500 | 5000 | 1 | * | 7.5 | 22 | 2 | * | 18 | * | 1500 | – | – | – | – | – | – | – | – | – |
| 13 | 40% | 500 | 4000 | 1 | * | 12 | 22 | 0 | * | 12 | * | 1270 | 1170 | 570 | 310 | 50 | 2000 | 1500 | 1000 | 500 | 600 |
| 14 | 50% | 500 | 4000 | 1 | * | 12 | 25.5 | 2 | * | 13 | 2 | 577 | 583 | 587 | 460 | 100 | 410 | 352 | 246 | 157 | 500 |
| 15 | 50% | 500 | 4000 | 1 | * | 12 | 25.5 | 2 | * | 18 | 2 | 656 | 664 | 681 | 450 | 80 | 4100 | 3500 | 2340 | 1877 | 1900 |
| 16 | Ref | 500 | 4000 | 1 | * | 12 | 25.5 | 2 | * | 25 | 2 | 110 | 50 | 30 | 100 | - | 1200 | 1600 | 2000 | 1700 | - |
| 17 | Ref | 500 | 4000 | 1 | * | 12 | 25.5 | 2 | * | 18 | 2 | 225 | 100 | 40 | 24 | 22 | 1100 | 1500 | 1200 | 1000 | - |

" * ", This step is not conducted during the fabrication.

" – ", This thickness is refused to be measured due to a micro-sized thickness of the channel 5.

The thickness of channel 5 (i.e. regardless of the dilution volume ratio) for sample #7-13 in Table 2 exceeds 1000 nm, and thus the corresponding parameters of UV photolithography are not suitable for fabrication of micro-channels with nanometer thickness. Furthermore, the comparison between the parameters of UV photolithography also reemphasizes the importance of the second step of soft- and post-exposure baking. These steps as shown in Table 2 were not carried out, intentionally, to study the variation of thickness under different parameters of the UV photolithography. As stated above, the second step of the soft baking at the higher temperature (~ 95° C) would lead to more evaporation of the solvent from the coated SU-8, and the second step of the post exposure baking (again at ~ 95° C) converts more monomers of the SU-8 through the cross-linking process into polymers as the result of the combination of these two steps significantly affect the change of SU-8 polymer from liquid state to solid state.

Considering sample #1-6 and #14-17 presented in Table 2, and also the situation where the photoresist with different dilution volume ratio integrated with all the processing steps required for a complete normal photolithography including spin coating, soft bake steps 1 and 2, exposure time, exposure dosage, post exposure bake steps 1 and 2 and developing under constant agitation over a controlled duration, sub-micrometer thickness SU-8 pattern samples were obtained. The micrographs of fabricated micro-channel with nanometer thickness are shown in Figures 2a-12a. Correspondingly, the related plots of surface profiling measured using Dektak surface profiler are illustrated in Figures 2b-12b. Taking into account Figures 2-7 and their corresponding process parameters in Table 2, the viscosity of the 20% - 25% diluted photoresists is a little thinner than the reference SU-8. Basically, the molecular structure of the diluted SU-8 has more space to move as a result of interaction between slow moving molecules of the reference SU-8 (i.e. undiluted SU-8) and fast moving of the solvent [21, 22]. Higher solvent volume ratio means less imposing force by the slow moving polymer molecules to the bounded fast moving solvent molecules.

As such, the combination of different speeds movement of the molecules to which rate of the solvent or polymer are mixed would produce different molecular structural of micro-channel and consequently different optical properties. Apart from this, this study aims to survey the effect of dilution along with photolithography parameters on the thickness of the micro-channels [23]. The influence of volume ratio of the solvent is evident from the two groups as shown by sample #1-6 and #14-17 in Table 2. Results are emphasizing that adding 50% or less of the solvent to the SU-8 polymer eventually produced the nanometer thickness micro-channels with a rectangular profile, however, assuming the SU-8 resist with the concentration above than 50% solvent also provides the nano-thickness channel but the quality of each channel reduce to lower grade. For distinguishing the differences between the produced channels based on the different concentration of the solvent in this study, the evaluation of the quality and thickness of the micro-channels is made by the use of the microscopic image and surface profiler, respectively. All the achieved results as defined in sample #1-6 and corresponding to Figures 2a-7a shows a rather qualified fabrication of the micro-channels. The thickness of each channel as can be seen from Figures 2b-7b is less than half a micron. Also, the slope of the channel's side walls is low enough to assume it as a rectangular hollow micro-channel. Obviously, the wider channels allow the more interaction of the UV light with the bed of the channel and thereby the more depth mainly can be observed for the channel 5 (i.e. the maximum resolution that can be achieved by the current probe of the surface profiler is 1 μm in diameter).

The thickness of channel 1 corresponding to all the situations shown in Figures 2-12 has the lowest thickness compare to the other channels (i.e. channels 2, 3, 4 and 5). The less contact of the UV light with the low width channel is the main reason for this observation (i.e. the more width of the channel the more intensity of the light will propagate through the aperture of the channel). However, the distance of UV light source from the substrate plane is known as another effective factor on the depth of the channels. For the conducted experiments in this study, the height of UV source from the plane was adjusted to the same distance and to deliver the maximum amount of the power to sample holder and photo mask. The thinnest thickness channels among the fabricated channels are obtained after conducting the test number 4 according to Table 2.

The outstanding factor which obviously causes the thickness of channel 5 to attain the value of 62 nm is changing the spin coater recipe so that the thickness of the SU-8 sample reaches the value of fewer than 1 μm. Examining the same procedure for the sample #8-12 without applying the pre- and post-baking never give a thickness in nanometer scale. As such, the effect due to the amalgamation of coating recipe with pre- and post- baking causes the most uniform channel thickness has been observed according to test number 4 Table 2. The less thickness of the coated SU-8 the better interaction of the light with channels even at lower width, viz. even at 1 μm channel width can be achieved.

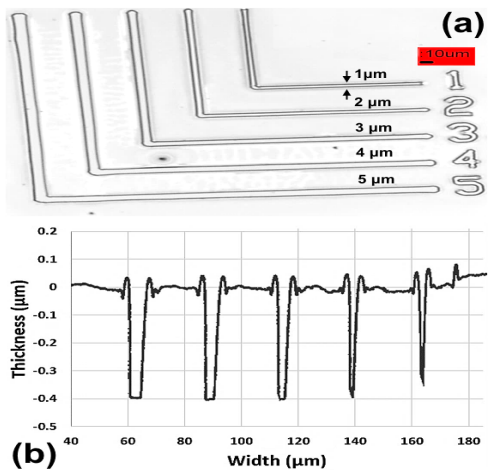

Fig. 2. Microscopic image of micro-channels with different width channels corresponding to the parameters of fabrication presented in test number 1 of table 2(a), the measured thickness of each channel by the use of surface profiler (b).

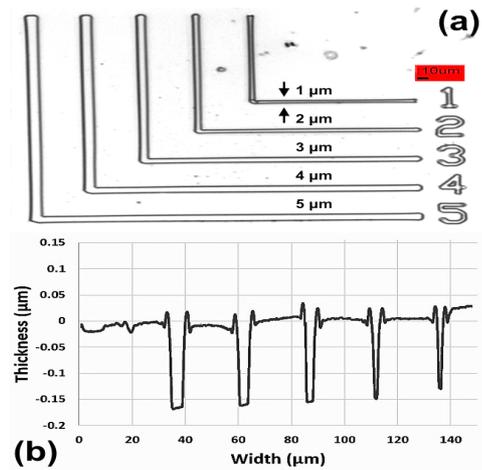

Fig. 3. Microscopic image of micro-channels with different width channels corresponding to the parameters of fabrication presented in test number 2 of Table 2(a), the measured thickness of each channel by the use of surface profiler (b).

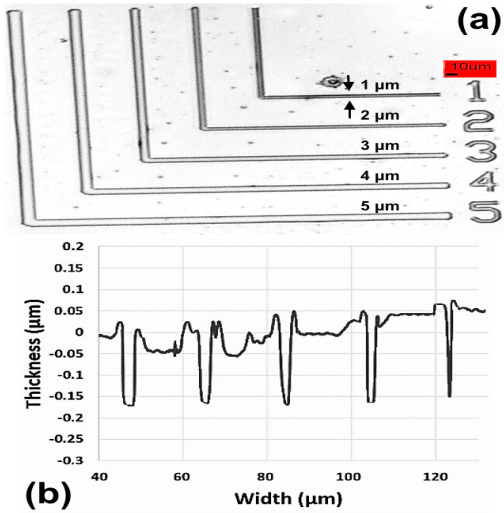

**Fig. 4.** Microscopic image of micro-channels with different width channels corresponding to the parameters of fabrication presented in test number 3 of Table 1(a), the measured thickness of each channel by the use of surface profiler (b).

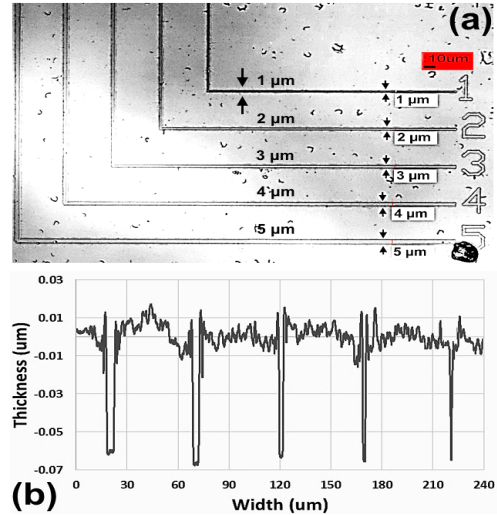

**Fig. 5.** Microscopic image of micro-channels with different width channels corresponding to the parameters of fabrication presented in test number 4 of Table 1(a), the measured thickness of each channel by the use of surface profiler (b).

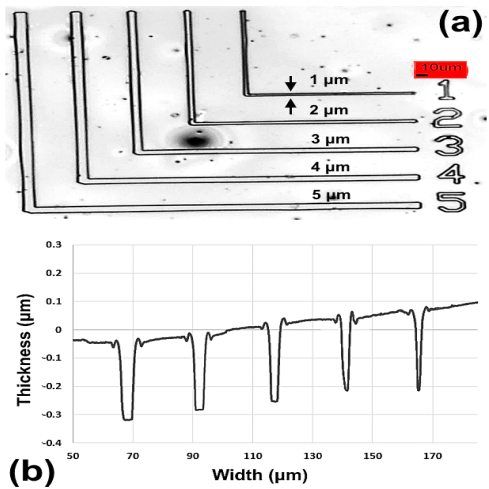

**Fig. 6.** Microscopic image of micro-channels with different width channels corresponding to the parameters of fabrication presented in test number 5 of Table 1(a), the measured thickness of each channel by the use of surface profiler (b).

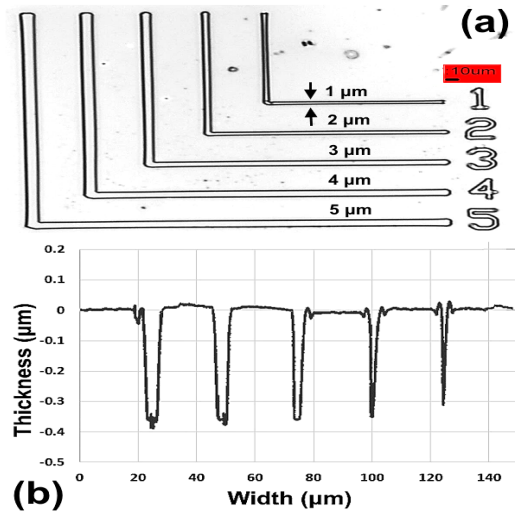

**Fig. 7.** Microscopic image of micro-channels with different width channels corresponding to the parameters of fabrication presented in test number 6 of Table 1(a), the measured thickness of each channel by the use of surface profiler (b).

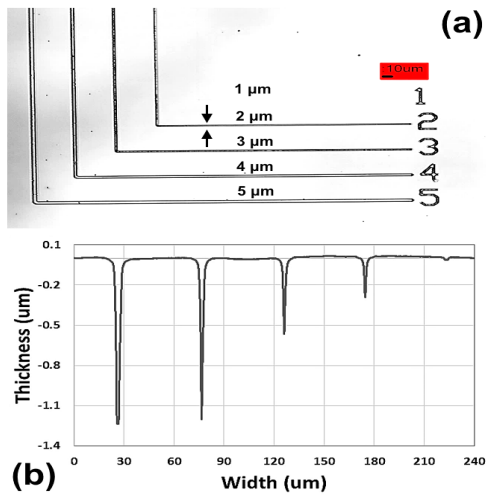

**Fig. 8.** Microscopic image of micro-channels with different width channels corresponding to the parameters of fabrication presented in test number 13 of Table 1(a), the measured thickness of each channel by the use of surface profiler (b).

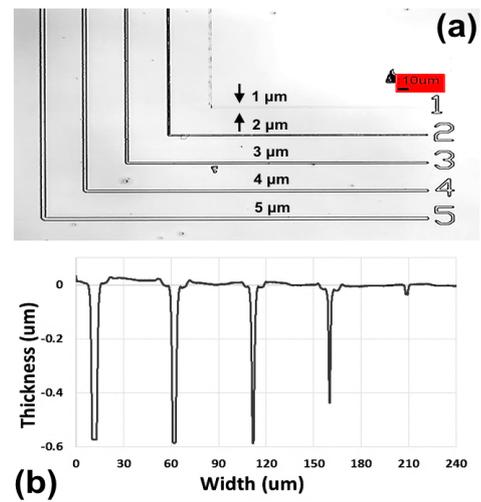

**Fig. 9.** Microscopic image of micro-channels with different width channels corresponding to the parameters of fabrication presented in test number 14 of Table 1(a), the measured thickness of each channel by the use of surface profiler (b).

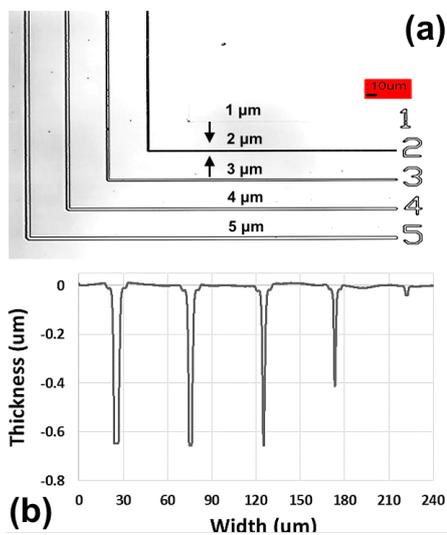

**Fig. 10.** Microscopic image of micro-channels with different width channels corresponding to the parameters of fabrication presented in test number 15 of Table 1(a), the measured thickness of each channel by the use of surface profiler (b).

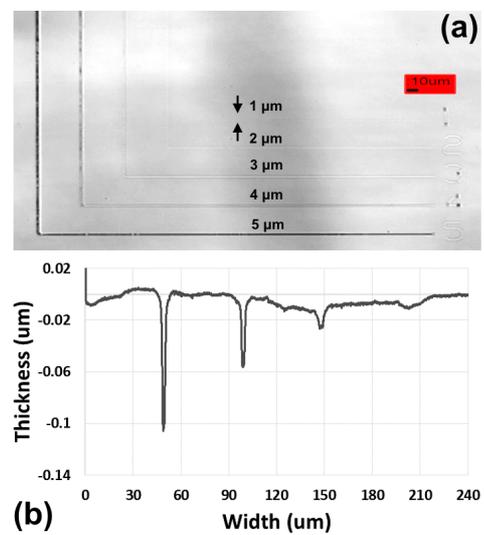

**Fig. 11.** Microscopic image of micro-channels with different width channels corresponding to the parameters of fabrication presented in test number 16 of Table 1(a), the measured thickness of each channel by the use of surface profiler (b).

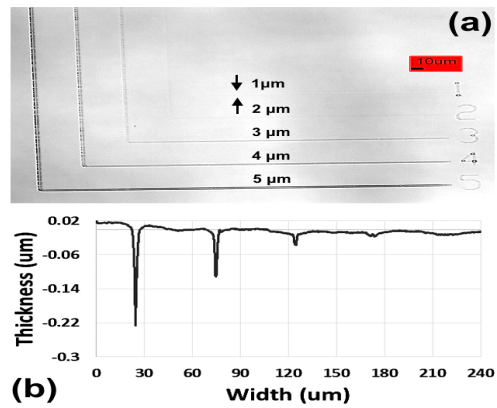

**Fig. 12.** Microscopic image of micro-channels with different width channels corresponding to the parameters of fabrication presented in test number 17 of Table 1(a), the measured thickness of each channel by the use of surface profiler (b).

Regardless the width of the channel and considering the effect of spin coater speed on the thickness of the channels formed by 20% diluted SU-8 samples, it can be seen from Figure 13 that there is an almost linear relation with a negative slope between the nano scale thickness and speed of the coater.

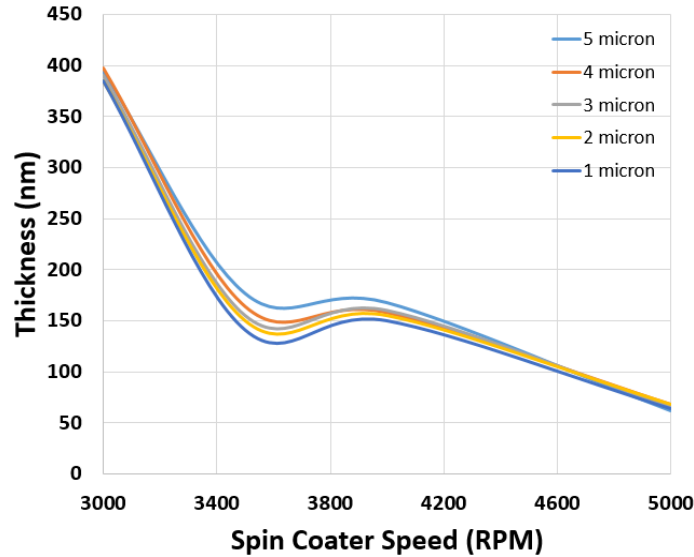

**Fig. 13.** The effect of spin coater speed on the thickness of the channels fabricated using 20% diluted SU-8 (2002)

The further evaluation of the results based on a combination of the different volume ratio of the diluted SU-8 and spin coater recipe also emphasize on the significant alteration from normal channel to very low-quality channel. For instance, comparing the thickness and quality of the channels in Figures 9 and 10 reveals that the highly diluted SU-8 viz. about 50%, is not suitable for nanometer thickness specifically when the small channel width (i.e. 1μm) is the aim of fabrication. This is perceptible as the original SU-8 can be used to fabricate micro-channel with nanometer thickness as to measurements shown in Figures 11 and 12, but the fabricated channels even at a longer developing time never show a standard feature of a rectangular channel.

## 3. Conclusion

From the aforementioned results, the inference can be drawn that the critical parameters of a standard photo lithography including the percentage of dilute photoresist SU-8, spin coating recipes, pre- and post-soft baking time as well as the temperature point and development time significantly alter the quality and thickness of the micro-channels. The results of this study suggest that the nano-thickness channels with acceptable rectangular shape can be fabricated by applying 20% diluted SU-8 coated over the silicon substrate and coating it with spinning at the speed of 5000 rpm, pre- and post-soft baking steps of 65°c and 95°c, over the duration of one minute for each step of baking, exposing of the photomask with power dosage of 25.5 mW /cm$^2$ with span time of 12 seconds and finally developing the sample by the SU-8 solvent over 12 seconds at the constant vibration with frequency of 2 MHz. The ability to control photolithography accuractly enables a varity of novel nanoph integrated nanophotonic devices and circuits to include light sources and emitters [24-26], reconfigurable switches and modulators [27-34], surface-sensitive devices [35], and application in on-chip data communication [36,37] ideally with nanoscale footprint for high electro-optic device efficiencies [38,39]

**Acknowledgements**

V.J.S. is supported by the U.S. Air Force Office of Scientific Research-Young Investigator Program under grant FA9550-14-1-0215, and under grant FA9550-14-1-0378 and FA9550-15-1-0447.

Reports, 6, 37419 (2016).